\documentclass[12pt]{iopart}
\usepackage{amsfonts}
\usepackage{graphicx}

\begin{document}

\title{Spectral methods for the detection of network community structure: A comparative analysis}

\author{Hua-Wei Shen and Xue-Qi Cheng}


\address{Institute of Computing Technology, Chinese Academy of Sciences,
Beijing, China}

\eads{\mailto{shenhuawei@software.ict.ac.cn,cxq@ict.ac.cn}}

%
%

\begin{abstract}
Spectral analysis has been successfully applied at the detection of
community structure of networks, respectively being based on the
adjacency matrix, the standard Laplacian matrix, the normalized
Laplacian matrix, the modularity matrix, the correlation matrix and
several other variants of these matrices. However, the comparison
between these spectral methods is less reported. More importantly,
it is still unclear which matrix is more appropriate for the
detection of community structure. This paper answers the question
through evaluating the effectiveness of these five matrices against
the benchmark networks with heterogeneous distributions of node
degree and community size. Test results demonstrate that the
normalized Laplacian matrix and the correlation matrix significantly
outperform the other three matrices at identifying the community
structure of networks. This indicates that it is crucial to take
into account the heterogeneous distribution of node degree when
using spectral analysis for the detection of community structure. In
addition, to our surprise, the modularity matrix exhibits very
similar performance to the adjacency matrix, which indicates that
the modularity matrix does not gain desired benefits from using the
configuration model as reference network with the consideration of
the node degree heterogeneity.
\end{abstract}


\maketitle

\section{Introduction}
Many complex systems in the real world can be modeled as graphs or
networks~\cite{Newman03a}. The topological characteristics and
dynamics on/of networks are critical to understanding the
relationship between structure and function of networks, such as the
modeling of networks~\cite{Cheng09}, the evolution of
networks~\cite{Zhang08,Zhang10}, the resilience of
network~\cite{Albert99,Cheng10a}, and capacity of
networks~\cite{Zhang07}. Many real world networks, including social
networks~\cite{Girvan02}, information networks~\cite{Flake02}, and
biological networks~\cite{Guimera05}, are found to divide naturally
into communities, known as groups of nodes such that the nodes
within a group are much more connected to each other than to the
rest of the network. Communities are of interest because they often
correspond to functional units such as the collections of pages on a
single topic on the Web~\cite{Flake02} and the pathways for
metabolic networks~\cite{Guimera05}.

The identification of community structure has attracted much
attention in various scientific fields. Many methods have been
proposed and applied successfully to some specific complex
networks~\cite{Newman04,Clauset04,Radicchi04,Palla05,Duch05,Newman06a,
Newman06b,Rosvall07,Bagrow08,Shen09a,Shen09b}. For reviews, the
reader can refer to~\cite{Fortunato10}. These methods are from
different perspectives, such as the centrality measures, link
density, percolation theory, and network compression. Besides these
methods, the spectral analysis has gained great success at
uncovering the community structure, respectively being based on the
adjacency matrix~\cite{Chauhan09}, the standard Laplacian
matrix~\cite{Arenas06}, the normalized Laplacian
matrix~\cite{Cheng10b}, the modularity
matrix~\cite{Newman06a,Newman06b}, the correlation
matrix~\cite{Shen10} and several other variants of these matrices.
However, to our knowledge, the comparison between these different
spectral methods is less reported. More importantly, it is still
unclear which matrix is more appropriate for the detection of
community structure.

In this paper, we conduct a comparative analysis of the
aforementioned five matrices on the benchmark networks which have
heterogeneous distributions of node degree and community size. The
comparison is carried out from two perspectives. The former one
focuses on whether the number of intrinsic communities can be
exactly identified according to the spectrum of these five matrices.
The latter evaluates the effectiveness of these matrices at
identifying the intrinsic community structure using their
eigenvectors. Test results show that the normalized Laplacian matrix
and the correlation matrix significantly outperform the other three
matrices. The possible reason is that these two matrices are both
normalized using the degree of nodes. Thus we can conclude that it
is crucial to take into account the heterogeneous distribution of
node degree when using spectral analysis for the detection of
community structure. In addition, to our surprise, the modularity
matrix exhibits very similar performance to the adjacency matrix,
which indicates that the modularity matrix does not gain desired
benefits from using the configuration model as reference network
with the consideration of the node degree heterogeneity.

Several comparative analyses on the methods for community detection
have been conducted~\cite{Danon05,Lancichinetti09}. Different from
them, our comparative analysis focuses on the matrices underlying
the spectral methods for community detection rather than comparing
the specific implementation of the existing spectral methods with
different heuristics to improve the final performance. This
comparison is fair and meaningful since the performance of spectral
methods heavily relies on the characteristics of the underlying
matrices.

\section{The matrices for spectral analysis}
\label{sec2}

The topology of network is often described in terms of adjacency
matrix. Based on the adjacency matrix, several other matrices are
formulated to investigate the properties of network, including the
standard Laplacian matrix, the normalized Laplacian matrix, the
modularity matrix and the correlation matrix. Existing studies
indicate that the spectrum of these matrices sheds light on the
community structure of network. In the following, we first give the
definition of these matrices and briefly introduce the methods to
detect the community structure using the spectrum of these matrices.

\begin{itemize}
  \item \emph{Adjacency matrix}. The elements $A_{ij}$ of an adjacency
  matrix $A$
  denote the strength of the edge connecting the nodes $i$ and $j$ if
  such an edge exists, and $0$ otherwise.
  (We restrict our attention in this paper to undirected networks.)
  In~\cite{Chauhan09}, the authors proposed that the spectrum
  of the adjacency matrix can unravel the number of communities.
  Specifically, the eigenvalues of the adjacency matrix is ranked in
  descending order, i.e., $\lambda_{1}^{A}\geq\lambda_{2}^{A}\geq\cdots
  \geq\lambda_{i}^{A}\geq\cdots\geq\lambda_{n}^{A}$, where $n$ is the
  number of network nodes. Each two successive eigenvalues form an
  eigengap, the $i$th eigengap being between
  $\lambda_{i}^{A}$ and $\lambda_{i+1}^{A}$ $(1\leq i \leq n-1)$. The
  length of the $i$th eigengap is $\lambda_{i}^{A}-\lambda_{i+1}^{A}$.
  Then, the number of communities is indicated by the place of the largest
  eigengap, i.e., $i$ is the number of communities if the largest
  eigengap is the $i$th one.

  \item \emph{Standard Laplacian matrix}. The standard Laplacian matrix is defined as
  $L=D-A$, where $D$ is a diagonal matrix with the diagonal element $D_{ii}$
  being the degree of the node $i$. As to the standard Laplacian matrix, the Fiedler's
  vector~\cite{Fiedler73,Fiedler75} has been well studied and widely used
  for two-way network partition. Fiedler's vector is the eigenvector of
  the standard Laplacian matrix corresponding to the second smallest eigenvalue.
  More importantly, the standard Laplacian matrix is often used to
  characterize the synchronization dynamics on networks~\cite{Arenas06,Yan09}. In~\cite{Arenas06},
  Arenas \textit{et al.} pointed out
  that the spectrum of the standard Laplacian matrix reveals the intrinsic
  topological scales. The eigenvalues are ranked in ascending order
  and the length of the $i$th eigengap is defined as
  $\log\lambda_{i+1}^{L}-\log\lambda_{i}^{L},(2\leq i \leq n-1)$\footnote{In this paper,
  we also tested the alternative eigengap defined as $\lambda_{i+1}^{L}-\lambda_{i}^{L},(1\leq i \leq n-1)$,
  and the results are similar.}. Then,
  $i$ is viewed as the appropriate candidate for the number of intrinsic
  communities if the $i$th eigengap is largest.

  \item \emph{Normalized Laplacian matrix}. The normalized Laplacian matrix
  is often defined as $N=I-T$, where $I$ is the identity matrix and $T$ is the
  transition matrix, which is defined as $T = D^{-1}A$ with the elements
  $T_{ij}$ being the probability that a random walker moves to the node $j$
  from the node $i$. The normalized Laplacian matrix is named after the fact that
  it can be written in the form $N=D^{-1}L$, i.e., normalizing the standard
  Laplacian matrix with the diagonal matrix $D$ of node degrees. In~\cite{Capocci05},
  the authors claimed that the spectrum of the transition matrix $T$ can be used to detect the community
  structure of networks. Actually, if $\lambda$ is an eigenvalue of the
  transition matrix, $1-\lambda$ is an eigenvalue of the normalized
  Laplacian matrix with the same eigenvector. Furthermore, the normalized
  Laplacian matrix is closely correlated to the diffusion dynamics on
  networks. Through investigating the diffusion dynamics on networks,
  Cheng and Shen pointed out~\cite{Cheng10b} that the community structure
  can be identified through the eigenvalues and eigenvectors of the
  normalized Laplacian matrix. Specifically, the eigenvalues are ranked
  in ascending order and the length of the $i$th eigengap is defined as
  $\lambda_{i+1}^{N}-\lambda_{i}^{N},(1\leq i \leq n-1)$. Then, $i$
  is viewed as the appropriate candidate for the number of intrinsic
  communities if the $i$th eigengap is largest.

  \item \emph{Modularity matrix}. The modularity matrix is proposed
  by Newman as a spectral explanation for the well-known measure,
  namely modularity, for the quality of network partition~\cite{Newman06a,Newman06b}.
  Its elements are defined as
  \begin{equation}
  B_{ij} = A_{ij} - \frac{k_{i}k_{j}}{2m},\nonumber
  \end{equation}
  where $k_{i}=\sum_{j}{A_{ij}}$ is the strength of the node $i$ and
  $2m=\sum_{ij}{A_{ij}}=\sum_{i}{k_i}$ is the total strength of all
  the nodes. In~\cite{Newman06b}, the eigenvectors corresponding to
  positive eigenvalues are utilized to uncover the community
  structure of networks. The number of communities can be determined
  according to the magnitude of the positive eigenvalues. Here, we rank
  the eigenvalues in descending order and the length of the $i$th eigengap
  is defined as $\lambda_{i-1}^{B}-\lambda_{i}^{B},(2\leq i\leq n)$. Then,
  $i$ is taken as the number of communities if the $i$th eigengap has the
  largest length. Note that, for the purpose of the detection of community
  structure, only the eigengaps among positive eigenvalues are considered.
  If all the eigenvalues are negative, no natural community structure exists,
  i.e., all the nodes belong to a sole community and the community number
  is $1$. In~\cite{Shen10}, the modularity
  matrix is shown to be the biased covariance matrix of network and the
  spectrum of the covariance matrix is investigated for the detection of the
  multiscale community structure.

  \item \emph{Correlation matrix}. The correlation matrix of network
  characterizes the correlation coefficients between pairs of nodes. Its
  element $C_{ij}$ are defined as
  \begin{equation}
  C_{ij}=\frac{B_{ij}}{\sqrt{k_{i}-k_{i}^{2}/2m}\sqrt{k_{j}-k_{j}^{2}/2m}}.\nonumber
  \end{equation}
  In~\cite{Shen10}, the correlation matrix is used to uncover the
  multiscale community structure of networks. Specifically, the eigenvalues are
  ranked in descending order and the length of the $i$th eigengap is
  defined as $\lambda_{i-1}^{C}-\lambda_{i}^{C},(2\leq i\leq n)$. The
  same to the modularity matrix, only the eigengaps among positive
  eigenvalues are considered. Then, $i$ is taken as the number of
  communities if the $i$th eigengap has the largest length. A similar
  matrix is called the symmetric normalized Laplacian matrix, whose
  element at the place $(i,j)$ is defined as
  $\delta_{ij}-A_{ij}/\sqrt{k_{i}}\sqrt{k_{j}}$, where $\delta_{ij}$ is $1$
  when $i=j$ and $0$ otherwise. This matrix is often used in
  spectral clustering algorithms together with the two
  aforementioned Laplacian matrices~\cite{Luxburg08}.
\end{itemize}

In summary, the number of communities can be determined according to
the eigengaps of the aforementioned five matrices. Actually, the
community structure can be further identified using the eigenvectors
of these matrices. Generally speaking, only several eigenvectors are
utilized to project each node into a low-dimensional node vectors,
and then the community structure is identified through clustering
the node vectors using, for example, the $k$-means clustering
method. Specifically, the selected eigenvectors correspond to the
largest $n_{c}$ eigenvalues for the adjacency matrix, the smallest
$n_{c}$ eigenvalues for the standard Laplacian matrix and the
normalized Laplacian matrix, the largest $n_{c}-1$ eigenvalues for
the modularity matrix and the correlation matrix. Here, $n_c$ is the
number of communities. These selected eigenvectors are stacked as
columns of a matrix and the transpose of the $i$th row of this
matrix is taken as the projected node vector corresponding to the
node $i$. The community structure is then detected through
clustering the projected node vectors.

Before proceeding, we first clarify why we choose the general method
for community detection using the $k$-means clustering method. On
one hand, this paper only considers the performance of the
aforementioned five matrices rather than the specific implementation
of spectral methods. Thus, it is fair and reasonable to choose the
general method, which determines the community number according to
the spectrum of matrices and identifies the community structure
using the eigenvectors of the matrices. On the other hand, the
$k$-means clustering method is the common practice for the spectral
clustering~\cite{Luxburg08}. Furthermore, the $k$-means clustering
method is facilitated by the projected node vector subspace spanned
by the top eigenvectors of matrices~\cite{Ding2003}. Thus, using the
$k$-means clustering on the node vectors provides a competitive
candidate among all the spectral methods for the detection of
community structure.

\begin{figure}
\hspace{70pt}\includegraphics[width=0.6\textwidth]{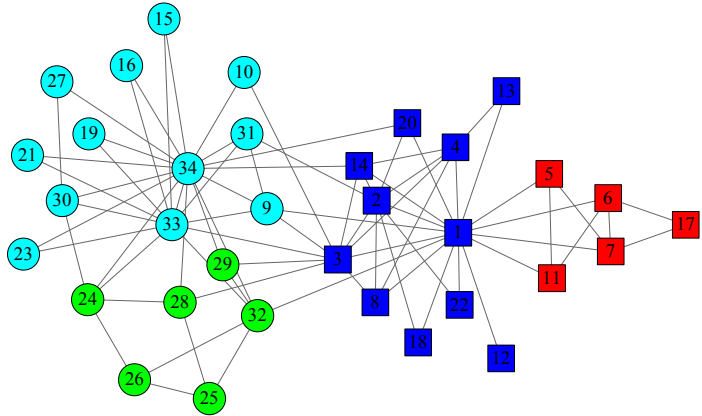}
\caption{The network of the karate club studied by
Zachary~\cite{Zachary77}. The real social fission of this network is
represented by two different shapes, circle and square. Another
meaningful partition is often found as the results of many community
detection method. The corresponding four communities are
differentiated by colors.}\label{fig1}
\end{figure}

\begin{figure}
\hspace{65pt}\includegraphics[width=0.85\textwidth]{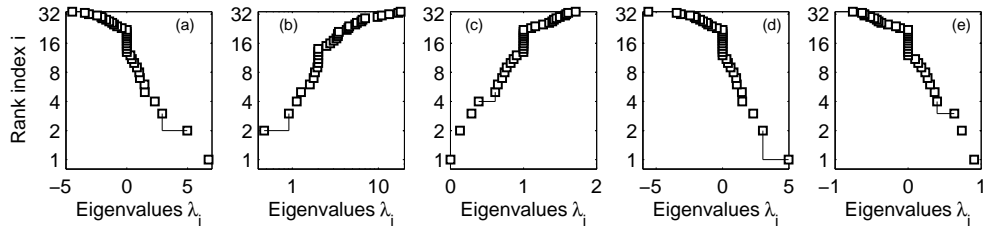}
\caption{The spectrum of the five considered matrices associated
with the Zachary's karate club network, respectively being (a) the
adjacency matrix, (b) the standard Laplacian matrix, (c) the
normalized Laplacian matrix, (d) the modularity matrix and (e) the
correlation matrix. For each matrix, the largest eigengap is marked
with an elbow line.}\label{fig2}
\end{figure}

Now, as an example, we illustrate the spectral methods through
application on the Zachary's karate club network, which has been
widely used to evaluate the community detection methods. This
network characterizes the social interactions between the
individuals in a karate club at an American university. A dispute
arose between the club's administrator and its principal karate
teacher, and as a result the club eventually split into two smaller
clubs, centered around the administrator and the teacher
respectively. The network and its fission is depicted in
figure~\ref{fig1}. The administrator and the teacher are represented
by nodes $1$ and $33$ respectively. Figure~\ref{fig2} shows the
spectrum of the aforementioned five matrices associated with the
Zachary's karate club network. The largest eigengap of the adjacency
matrix, the standard Laplacian matrix and the modularity matrix
indicate that the optimal number of community is $2$. The
corresponding community structure is consistent with the real split
of the network. However, as indicated by the largest eigengap of the
normalized Laplacian matrix and the correlation matrix, $4$ is the
optimal number of communities. The corresponding four communities
are shown in figure~\ref{fig1} differentiated with colors, which is
the results of many existing methods for community detection
including the modularity maximization.

\begin{figure}
\hspace{65pt}\includegraphics[width=0.85\textwidth]{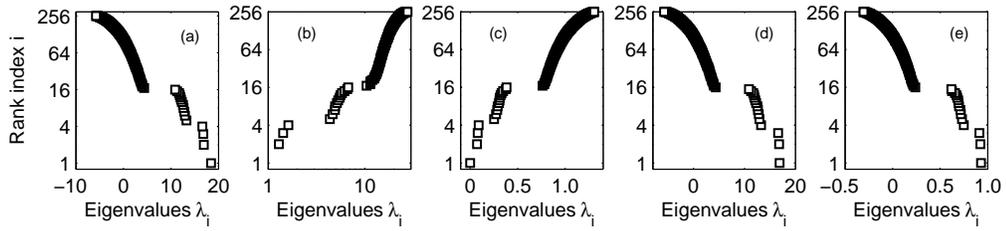}
\caption{The spectrum of the five matrices associated with the H13-4
network, respectively being (a) the adjacency matrix, (b) the
standard Laplacian matrix, (c) the normalized Laplacian matrix, (d)
the modularity matrix and (e) the correlation matrix.}\label{fig3}
\end{figure}

As illustrated by the previous example, the five matrices give rise
to two different resulting partitions as the community structure of
the network. Actually these two partitions correspond to two
different topological scales of the network. The multiple scale of
topological description is a common phenomenon in real-world
networks~\cite{Shen09a,Shen10,Fortunato07,Arenas08,Lambiotte08,Ronhovde09,Ahn2010}.
Actually, the multiscale community structure can be revealed through
considering more eigengaps besides the largest one among the
eigenvalues of the aforementioned five matrices. As an example, we
illustrate the detection of the multiscale community structure of
the H13-4 network, which is constructed according
to~\cite{Arenas06}. This network has two predefined hierarchical
levels. The first hierarchical level consists of $4$ groups of $64$
nodes and the second hierarchical level consists of $16$ groups of
$16$ nodes. On average, each node has $13$ edges connecting to the
nodes in the same group at the second hierarchical level and has $4$
edges connecting to the nodes in the same group at the first
hierarchical level. This explains the name of such kind of networks.
In addition, the average degree of each node is $18$. According to
the construction rules of the H13-4 network, the two hierarchical
levels constitute the different topological descriptions of the
community structure of the H13-4 network at different scales. As
shown in figure~\ref{fig3}, the community numbers associated with
the two predefined topological scales are clearly revealed by the
top two largest eigengaps occurring in the spectrum of the five
matrices. The resulting communities are exactly the predefined node
groups under the two hierarchical levels. However, according to the
length of eigengap, the standard Laplacian matrix seems to prefer
the first hierarchical level while the other four matrices tend to
reveal the second hierarchical level.

\begin{figure}
\hspace{65pt}\includegraphics[width=0.85\textwidth]{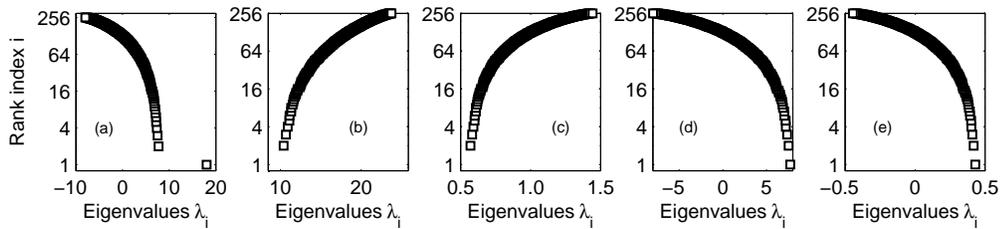}
\caption{The spectrum of the five matrices associated with the
randomized H13-4 network, respectively being (a) the adjacency
matrix, (b) the standard Laplacian matrix, (c) the normalized
Laplacian matrix, (d) the modularity matrix and (e) the correlation
matrix.}\label{fig4}
\end{figure}

Furthermore, we apply all these matrices to the random network. For
comparison, we construct the random network through shuffling the
edges of the H13-4 network. Figure~\ref{fig4} shows the spectrum of
the five matrices associated with the randomized H13-4 network. The
spectrum of these matrices indicate that the number of communities
is $1$ or $256$, i.e., all the nodes belong to the same community or
each node forms a community. This findings are reasonable since it
is commonly believed that randomized networks have no community
structure.

\begin{figure}
\hspace{65pt}\includegraphics[width=0.4\textwidth]{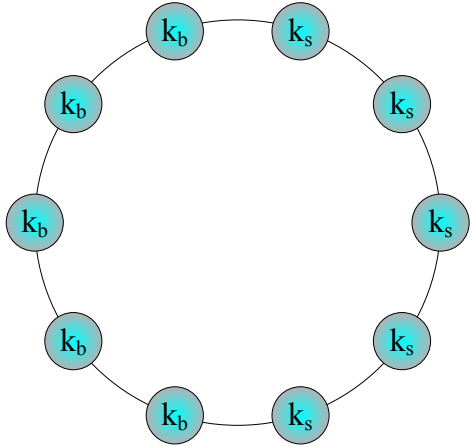}
\caption{The clique circle network as a schematic example. Each
circle corresponds to a clique, whose size is marked by its label
$k_{s}$ or $k_{b}$. }\label{fig5}
\end{figure}

\begin{figure}
\hspace{65pt}\includegraphics[width=0.85\textwidth]{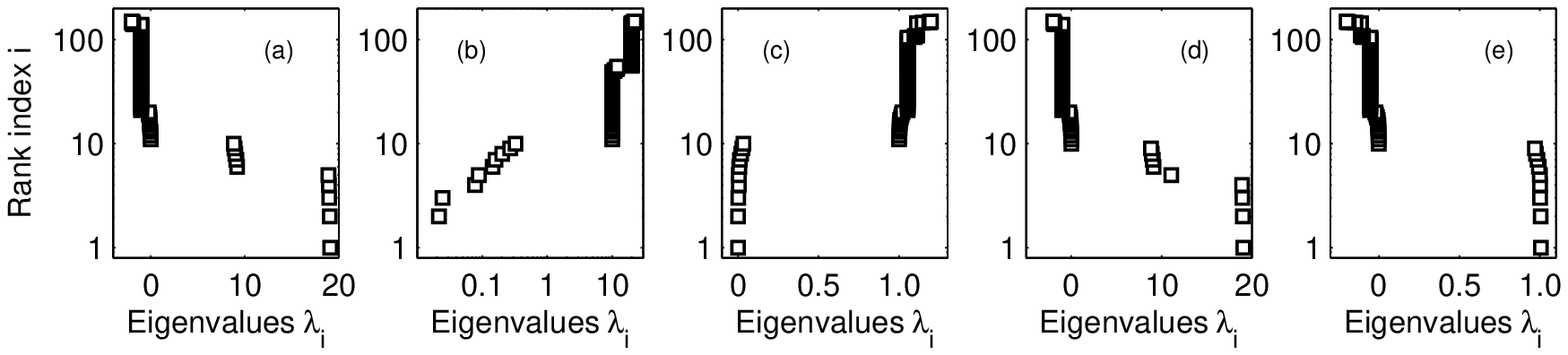}
\caption{The spectrum of the five matrices on the clique circle
network with the clique size being $k_{s}=10$ and $k_{b}=20$. These
matrices are respectively (a) the adjacency matrix, (b) the standard
Laplacian matrix, (c) the normalized Laplacian matrix, (d) the
modularity matrix and (e) the correlation matrix.}\label{fig6}
\end{figure}

\begin{figure}
\hspace{65pt}\includegraphics[width=0.85\textwidth]{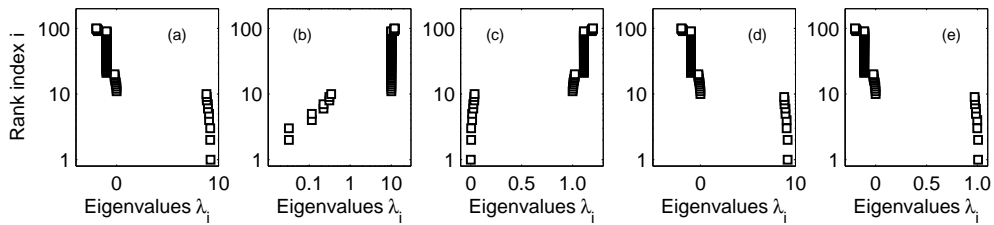}
\caption{The spectrum of the five matrices on the clique circle
network with the clique size being $k_{s}=k_{b}=10$. These matrices
are respectively (a) the adjacency matrix, (b) the standard
Laplacian matrix, (c) the normalized Laplacian matrix, (d) the
modularity matrix and (e) the correlation matrix.}\label{fig7}
\end{figure}

The previous examples show that the aforementioned five matrices are
both effective at revealing the community structure of network. Note
that, as to the example H13-4 network, the nodes have approximately
the same degree and the communities at a specific scale are of the
same size. However, the real world networks usually have
heterogenous distributions of node degree and community size. Thus
it will be more convincing to test these matrices on networks with
heterogenous distributions of node degree and community size. Before
we give such a test in the subsequent section, using a schematic
network, we first illustrate the difference between the
effectiveness of these matrices. The schematic network is often
called the clique circle network as depicted in figure~\ref{fig5}.
Generally speaking, the intrinsic community structure corresponds to
the partition where each clique is taken as a community, which is
the sole intrinsic scale existing in this network. As shown in
figure~\ref{fig6}, the sole topological scale is exactly revealed by
the spectrum of the standard Laplacian matrix, the normalized
Laplacian matrix and the correlation matrix. However, two scales are
observed when we investigate the community structure of this network
using the spectrum of the adjacency matrix and the modularity
matrix. One scale corresponds to the intrinsic scale of the network,
and the other corresponds to the partition dividing the network
nodes into $5$ groups, which is not desired. In~\cite{Fortunato07},
Fortunato \textit{et al} pointed out the resolution limit problem of
the modularity through investigating the modularity maximization on
such a clique circle network with each clique having the same size.
However, as shown in figure~\ref{fig7}, when all the cliques have
the same size (i.e., the homogeneous node degree), the intrinsic
community structure can be exactly revealed by all the five
matrices, including the modularity matrix. This indicates that the
resolution limit problem of the modularity is not the same to the
problem studied in this paper. Specifically, the resolution limit
problem means that there exists an intrinsic scale beyond which the
smaller communities cannot be detected through maximizing the
modularity. As to the heterogeneity problem of the modularity matrix
considered in this paper, we focus on whether the modularity matrix
can reveal the natural community structure, which can be detected
using the spectral clustering method instead of the modularity
maximization. In sum, the resolution limit problem talks about the
maximization of modularity while the heterogeneity problem takes
root in the modularity matrix. Thus we claim that it is crucial to
deal with the heterogeneous degree when using the spectral methods
for community detection.

\section{Tests on benchmark networks}
\label{sec3}

In this section, we show the effectiveness of the aforementioned
five matrices at identifying the community structure on benchmark
networks. We utilize the benchmark proposed by Lancichinetti
\textit{et al.}~\cite{Lancichinetti08}. This benchmark provides
networks with heterogeneous distributions of node degree and
community size. Thus it poses a much more severe test to community
detection algorithms than Newman's standard
benchmark~\cite{Newman04}. Many parameters are used to control the
generated networks in this benchmark: the number of nodes $N$, the
average node degree $\langle k\rangle$, the maximum node degree
max$\rule[-1pt]{0.15cm}{0.3pt}k$, the mixing ratio $\mu$, the
exponent $\gamma$ of the power law distribution of node degree, the
exponent $\beta$ of the power law distribution of community size,
the minimum community size min$\rule[-1pt]{0.15cm}{0.3pt}c$, and the
maximum community size max$\rule[-1pt]{0.15cm}{0.3pt}c$. In our
tests, we use the default parameter configuration where $N=1000$,
$\langle k\rangle=15$, max$\rule[-1pt]{0.15cm}{0.3pt}k=50$,
min$\rule[-1pt]{0.15cm}{0.3pt}c=20$, and
max$\rule[-1pt]{0.15cm}{0.3pt}c=50$. To test the influence from the
distribution of node degree and community size, we adopt four
parameter configurations for $\gamma$ and $\beta$, respectively
being $(\gamma,\beta)=(2,1)$, $(\gamma,\beta)=(2,2)$,
$(\gamma,\beta)=(3,1)$ and $(\gamma,\beta)=(3,2)$. Finally, by
tuning the parameter $\mu$, we test the effectiveness of the five
matrices on the networks with different fuzziness of community
structure. The larger the mixing ratio parameter $\mu$, the fuzzier
the community structure of the generated network.

\begin{figure}
\hspace{65pt}\includegraphics[width=0.65\textwidth]{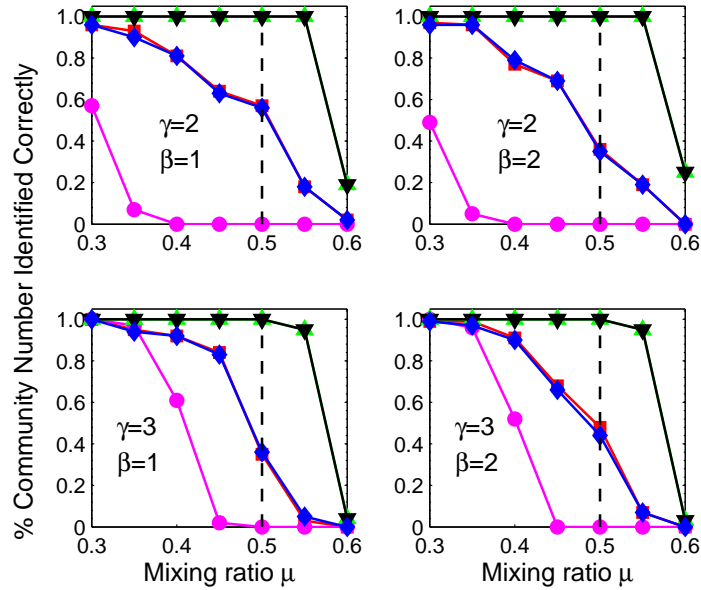}
\caption{The effectiveness of the five spectral methods at
identifying exactly the number of intrinsic communities on benchmark
networks with different parameter configurations. For each parameter
configuration, $100$ generated networks are used. The corresponding
matrices are respectively the adjacency matrix ($\Box$), the
standard Laplacian matrix ($\bigcirc$), the normalized Laplacian
matrix ($\bigtriangleup$), the modularity matrix ($\lozenge$) and
the correlation matrix ($\bigtriangledown$).} \label{fig8}
\end{figure}

The first test focuses on whether the number of communities can be
correctly identified. Note that each benchmark network has only one
significant topological scale according to the construction rules.
Thus we only consider whether such a scale can be revealed by the
largest eigengap in the spectrum of the five matrices. For each
given mixing ratio $\mu$, $100$ benchmark networks are generated.
For each network, we use the spectrum of the aforementioned five
matrices to identify the number of communities. The performance of
each method is characterized by the fraction of benchmark networks
whose community number is correctly identified. As shown in
figure~\ref{fig8}, the best results are obtained by the methods
based on the normalized Laplacian matrix and the correlation matrix,
which actually give the identical results for all the four used
parameter configurations. When the mixing ratio $\mu$ is smaller
than $0.5$, i.e., the communities are defined in the strong
sense~\cite{Radicchi04}, the number of communities can be accurately
identified by investigating the spectrum of the normalized Laplacian
matrix or the correlation matrix. Even when $\mu$ is larger than
$0.5$ (e.g., $0.55$), these two matrices still give rather good
results. The adjacency matrix and the modularity matrix exhibit
rather similar effectiveness, obtaining very good results when the
community structure is evident and deteriorating when the community
becomes fuzzier with the increase of the mixing ratio $\mu$.
Compared with the other four matrices, the standard Laplacian matrix
gives the worst results, failing to identify the correct number of
communities even when the community structure is quite evident. In
addition, the exponent $\gamma$ of the power law distribution of
node degree affects the effectiveness of the matrices except the
normalized Laplacian matrix and the correlation matrix. The possible
reason is that these two matrices take into account the distribution
of node degree through the normalization operation in their
definition. Finally, as shown in figure~\ref{fig8}, it seems that
all these five matrices are not very sensitive to the exponent
$\beta$ of the power law distribution of community size.

\begin{figure}
\hspace{65pt}\includegraphics[width=0.65\textwidth]{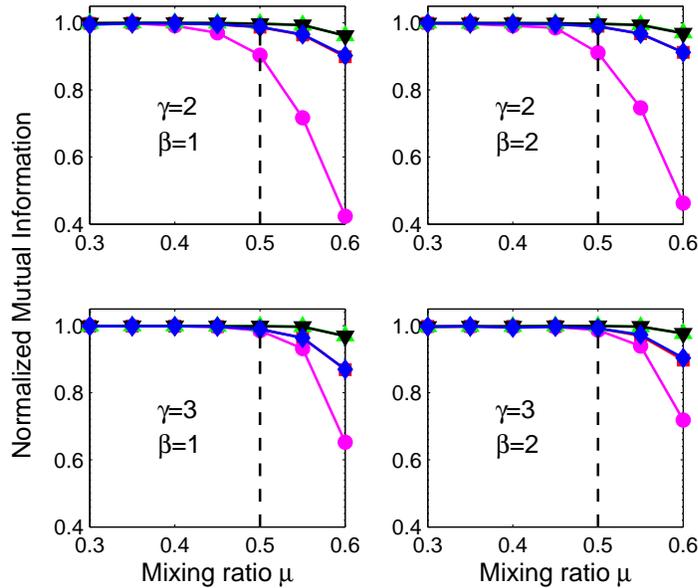}
\caption{The effectiveness of the five spectral methods at
identifying intrinsic community structure on benchmark networks with
different configuration parameters. Each point corresponds to an
average over $100$ network realizations for each parameter
configuration. The corresponding matrices are respectively the
adjacency matrix ($\Box$), the standard Laplacian matrix
($\bigcirc$), the normalized Laplacian matrix ($\bigtriangleup$),
the modularity matrix ($\lozenge$) and the correlation matrix
($\bigtriangledown$).}\label{fig9}
\end{figure}

The second test turns to the performance of the eigenvectors of the
five tested matrices. Given the number of communities, we
investigate whether the predefined community structure can be
identified using the eigenvectors of the five tested matrices. The
corresponding community detection methods cluster the projected node
vectors using the $k$-means clustering method. Each method produces
a network partition to represent the community structure. To compare
the partition found by these methods with the answer network
partition, we adopt the normalized mutual information
(NMI)~\cite{Danon05} to reflect the effectiveness of each method.
The larger the NMI, the more effective the tested method. As shown
in figure~\ref{fig9}, the same to the first test, the normalized
Laplacian matrix and the correlation matrix give the best and almost
identical results. The adjacency matrix and the modularity matrix
also exhibit the similar performance, being a little worse than the
normalized Laplacian matrix and the correlation matrix. As to the
standard Laplacian matrix with the worst performance, the NMI even
reaches $0.4$ when the mixing ratio $\mu$ is up to $0.6$ with
$\gamma=2$. Furthermore, the heterogeneous distribution of the node
degree affects the NMI of the spectral methods based on the
adjacency matrix, the modularity matrix and especially the standard
Laplacian matrix.

In summary, the normalized Laplacian matrix and the correlation
matrix outperforms the other three matrices both at identifying the
number of communities according to the spectrum and identifying the
community structure using the top eigenvectors. This indicates that
it is crucial to take into account the heterogeneous distribution of
node degree when using spectral analysis for the detection of
community structure. In addition, although the modularity considers
the heterogeneity through introducing the null-model reference
network (i.e., the configuration model), as shown in~\cite{Shen10},
this operation is in fact a kind of translation transformation and
thus cannot alleviate the influence on the detection of community
structure from the heterogeneous distribution of node degree. This
phenomenon can be seen from the experimental results on the
Lancichinetti's benchmark networks, i.e., the modularity matrix
obtains very similar results to the adjacency matrix.

\section{Conclusions}
\label{sec4}

We have carried out a comparative analysis on the spectral methods
for the detection of network community structure through evaluating
the performance of five widely used matrices on the benchmark
networks with heterogeneous distribution of node degree and
community size. These five matrices are respectively the adjacency
matrix, the standard Laplacian matrix, the normalized Laplacian
matrix, the modularity matrix and the correlation matrix. Test
results demonstrate that the normalized Laplacian matrix and the
correlation matrix significantly outperform the other three matrices
at identifying the community structure of networks. This indicates
that the heterogeneity of node degree is a crucial ingredient for
the detection of community structure using spectral methods and the
matrices that do not properly account for it are doomed to fail or
to give not very accurate results. In addition, to our surprise, the
modularity matrix does not gain desired benefits from using the
configuration model as reference network with the consideration of
the node degree heterogeneity.

\ack{This work was funded by the National Natural Science Foundation
of China under grants Nos $60873245$ and $60933005$, and the
National Basic Research Program of China ($973$) under grant No
$2007CB310805$. The authors gratefully acknowledge S Fortunato and A
Lancichinetti for providing the test benchmark. The authors thank
the anonymous reviewers for valuable comments on this paper. The
authors also thank J M Huang, P Du, X F Zhu, and P Cao for useful
discussions and suggestions.}

\end{document}